\documentstyle[aps,prc,twocolumn,epsfig]{revtex}
\begin{document}

\title{On the origin of order from random two-body interactions}


\author{S. Dro\.zd\.z and M. W\'ojcik}

\address{Institute of Nuclear Physics, PL--31-342 Krak\'ow, Poland} 
\date{\today}
\maketitle

\begin{abstract}
We investigate the origin of order in the low-lying spectra of many-body
systems with random two-body interactions. Contrary to the common belief
our study based both on analytical as well as on numerical arguments
shows that these are the higher $J$-sectors whose ground states are more orderly
than the ones in the $J=0$-sector. A predominance of $J=0$ ground states  
turns out to be the result of putting on together states with different 
characteristic energy scales from different $J$-sectors.  

\end{abstract}


\bigskip

The nature of mechanism generating order out of randomness constitutes
one of the most fundamental issues of the contemporary physics.  
Theories based on various versions of ensembles of the random matrices
provide one possible theoretical frame for studying such effects.
In this context the recently identified~\cite{Johnson} preponderance of the 
$J=0$ ground states in strongly interacting Fermi systems, such as atomic nuclei,
arising from random two-body interactions came as a surprise 
since there is no obvious pairing character in the assumed random force.  
Various possible explanations of this effect have been tested~\cite{Bijk}
with no success, however. One purpose of the present note to provide     
a consistent picture of the mechanism generating this effect.

Schematically, indicating nevertheless all the relevant ingredients,
the interaction matrix elements $v^J_{\alpha,\alpha'}$ of good total
angular momentum $J$ in the shell-model 
basis $\vert {\alpha} \rangle$ can be expressed as follows~\cite{Talmi}:
\begin{equation}
v^J_{\alpha,\alpha'} = \sum_{J'} \sum_{i i'} 
c^{J \alpha \alpha'}_{J' i i'}
g^{J'}_{i i'}.
\label{eqv}
\end{equation}   
The summation runs over all combinations of the two-particle states 
$\vert i \rangle$ coupled to the angular momentum $J'$ and connected 
by the two-body interaction $g$. 
$g^{J'}_{i i'}$
denote the radial parts of the corresponding two-body matrix elements while 
$c^{J \alpha \alpha'}_{J' i i'}$ 
globally represent elements 
of the angular momentum recoupling geometry. 

In statistical ensembles of matrices the crucial factor determining the
structure of eigenspectrum is the probability distribution $P_V(v)$ of matrix
elements~\cite{Drozdz}. Especially relevant are the tails of such distributions
since they prescribe the probability of appearance of the large matrix
elements. From the point of view of the mechanism producing the energy gaps
they are most effective in generating a local reduction of dimensionality
responsible for such effects. In principle, the probability distribution of
the shell model matrix elements is prescribed by their general structure 
expressed by the eq.~(\ref{eqV}), provided the probability distributions 
of both $g^{J'}_{i i'}$ and $c^{J \alpha \alpha'}_{J' i i'}$ are known.
In general terms this structure can be considered to have the following form:
\begin{equation}
V = V_1 + V_2 + ... + V_N
\label{eqV}
\end{equation} 
and each $V_i$ to be a product of another two variables denoted 
as $C_i$ and $G_i$. 
By making use of the convolution theorem~\cite{Bracewell} the probability 
distribution $P_V(v)$ 
that $V$ assumes a value equal to $v$ can be expressed as:
\begin{equation}
P_V (v) = 
F^{-1} [F(P_{V_1}(v_1)) \cdot F(P_{V_2}(v_2)) \cdot ... \cdot F(P_{V_N}(v_N))],   
\label{eqPV}
\end{equation}
where $F$ denotes a Fourier transform, $F^{-1}$ its inverse and 
$P_{V_i}(v_i)$ the probability distributions of individual terms 
in eq.~(\ref{eqV}). Taking in addition into account the fact that 
\begin{equation}
P_{V_i} (v_i) = \int dg_i P_{G_i} (g_i) P_{C_i} ({v_i \over g_i}) 
{1 \over {\vert g_i \vert}}
\label{eqPVi}
\end{equation}
one can explicitely derive the form of $P_V (v)$ in several cases.  
Assuming for instance that all the above constituents are identically
Gaussian distributed (then, according to eq.~(\ref{eqPVi}),
$P_{V_i}(v_i) = K_0 (\vert v_i \vert) /\pi$ and thus
$F(P_{V_i}(v_i)) = 1/\sqrt {1 + \omega^2}$ ) one arrives at
\begin{equation}
P_V (v) = { { \vert v \vert^{(N-1)/2} K_{(N-1)/2} (\vert v \vert) } \over
{ 2^{(N-1)/2} \Gamma(N/2) \sqrt{\pi} }},
\label{eqPr}
\end{equation}
where $K$ stands for the modified Bessel function.
Asymptotically, for large $v$, this leads to 
\begin{equation}
P_V(v) \sim \exp (-\vert v \vert)~{\vert v \vert}^{N/2-1}.
\label{eqPa}
\end{equation}

For such a global estimate the identical Gaussian  distribution of
$g^{J'}_{i i'}$
is consistent both with the Two-Body Random Ensemble (TBRE)~\cite{French}
and with the Random Quasiparticle Ensemble (RQE)~\cite{Johnson}.
The only anticipated difference originates from the fact that in the 
second case the variance of the distribution drops down with $J'$ like
the inverse of $2J' + 1$ which is expected to result in a smaller effective $N$
as compared to TBRE. By contrast, 
in both versions of the above random ensembles the geometry expressed by  
$c^{J \alpha \alpha'}_{J' i i'}$ 
enters explicitely. However, the complicated
quasi-random coupling of individual spins is believed~\cite{Ericson} to
result in the so-called geometric chaoticity~\cite{Zelev}. If applicable indeed
then this fact implies validity of the above estimate for $P_V(v)$.   
Since this is an important element for the present considerations below
we explicitely verify its range of applicability.   
 
The model to be quantitatively explored here consists, 
similarly as in ref.~\cite{Johnson}, of 6 identical particles 
(all single-particle energies are set to zero) operating in
the $sd$ shell. From the nuclear spectroscopy point of view this can be 
identified as $^{22}O$. Fig.~1 shows distributions of the corresponding
geometrical factors 
$c^{J \alpha \alpha'}_{J' i i'}$ 
for $\alpha \ne \alpha'$
and for the relevant values of $J$ and $J'$.\\ 
\vskip 0.3cm
\epsfig{file=fig1.eps,width=8cm}
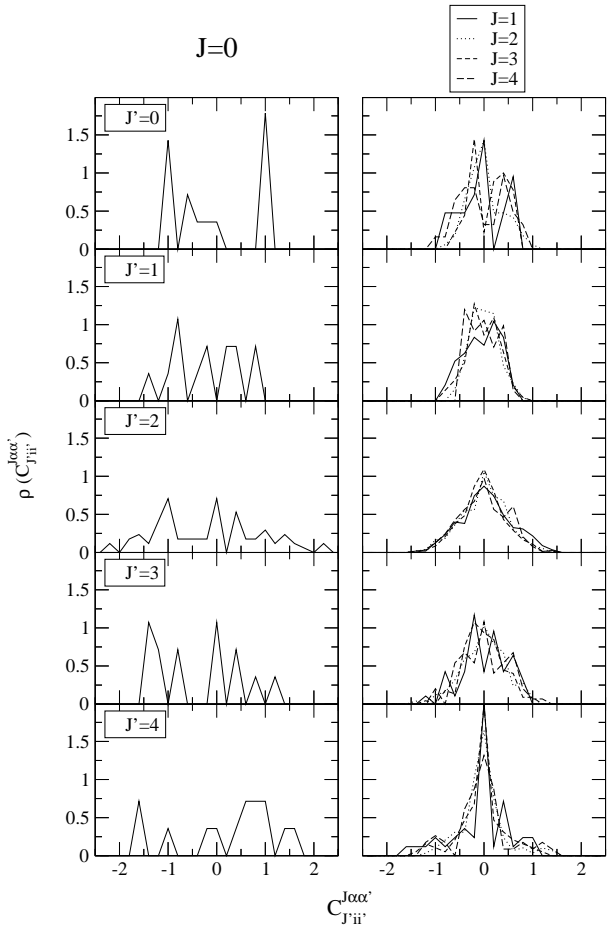
{\bf Fig.~1} The normalised distribution of geometrical factors  
$c^{J \alpha \alpha'}_{J' i i'}$ entering the off-diagonal matrix elements
(eq.~(\ref{eqv})) for the model of 6 particles in the $sd$-shell. \\
\vskip 0.3cm
As one can see, the Gaussian
may be considered quite a reasonable representation of the distribution of such
factors for all combinations of $J$ and $J'$ with one exception for those
which involve $J=0$. In this later case the distribution of    
$c^{0 \alpha \alpha'}_{J' i i'}$ resembles more a uniform distribution 
over a finite interval located symmetrically with respect to zero.
These empirical facts justify well the estimates of $P_V(v)$ based on 
eq.~(\ref{eqPr}) for $J \ne 0$ and not so well for $J=0$.   
More appropriate in this particular case is to assume a uniform distribution
of $c^{0 \alpha \alpha'}_{J' i i'}$ over an interval confined by say
$-c_0$ and $c_0$, i.e., $P_{C_i}(c_i) = 1/2c_0$, retaining $P_{G_i}(g_i)$
in its original Gaussian form of course.  
By making use of eqs.~(\ref{eqPV}) and (\ref{eqPVi}) one then obtains
\begin{equation}
P_V(v) = {1 \over {\sqrt {2\pi}}} \int_0^{\infty} 
\Bigl [ {{erf(c_0 \omega/\sqrt{2})} \over {c_0 \omega}} \Bigr ]^N
\cos (\omega v) d\omega
\label{eqPVu}
\end{equation}
which for large $v$ behaves like
\begin{equation}
P_V(v) \sim \exp(-{\vert v \vert}^2).
\label{eqPVul}
\end{equation}

An explicit calculation of the distribution of the shell model 
off-diagonal matrix elements for the various $J$-values based on the present
model with two-body matrix elements drawn from RQE (TBRE results in similar
relations among different $J$-sectors though the distributions are somewhat
broader as compared to RQE) confirms the above analytical estimates as is 
illustrated in Fig.~2.\\
\vskip 1.1cm
\epsfig{file=fig2.eps,width=8cm} 
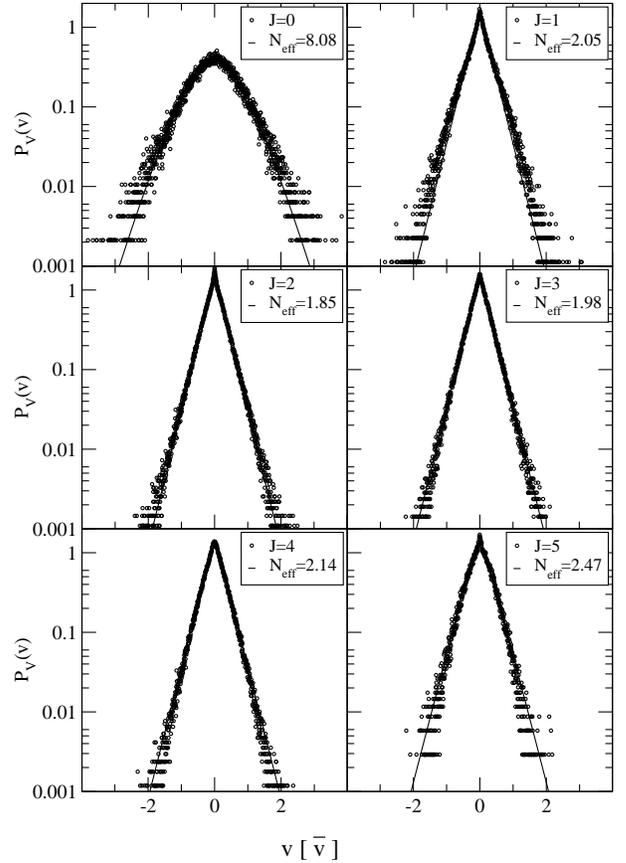
{\bf Fig.~2} The probability distribution of nonzero many-body off-diagonal
matrix elements in different $J$-sectors drawn from one thousand of RQE 
samples of two-body matrix elements. The energy scale is set by ${\bar v}$ 
where $w_{J'}={\bar v}^2 / (2J'+1)$ and $w_{J'}$ determines the RQE mean 
square variance. These distributions are fitted (continuous
solid lines) in terms of eq.~(\ref{eqPr}) with $N$ treated as a fitting 
parameter. The corresponding best $N$'s $(N_{eff})$ for each $J$ are listed.
By increasing $N$ the distribution prescribed by eq.~(\ref{eqPr})
quickly approaches (as a consequence of the central limit theorem) the Gaussian
distribution. In this way the $J=0$ distribution is demonstrated to be much
closer to the Gaussian than the remaining ones whose asymptotic behavior is 
consistent with a slower, exponential fall-off. \\      
\vskip 0.3cm
Indeed, such a distribution in the $J=0$ sector resembles more a Gaussian
and the tails of this distribution drop down faster as compared to the
$J \ne 0$-sectors where the large $v$ tails drop down slower, as consistent    
with an exponential asymptotics of eq.~(\ref{eqPV}). At the same time
the $J \ne 0$ sectors are dominated by very small matrix elements to a larger
degree than $J=0$. The probability of
appearance of a large off-diagonal matrix element which in magnitude
overwhelms the remaining ones is thus greater for $J \ne 0$ than for $J=0$.
Such an effective reduction of the rank in the former case is expected 
to result in a stronger tendency to localization as compared 
to GOE~\cite{Drozdz,Cizeau}. The corresponding characteristics can be 
quantified in terms of the information entropy
\begin{equation}
K_l^J = - \sum_{\alpha=1}^{M_J} 
\vert a_{l,\alpha}^J \vert^2  \ln \vert a_{l,\alpha}^J \vert^2    
\label{eqE}
\end{equation}
of an eigenstate labelled by $l$ from the $J$-sector. The coefficients
$a_{l,\alpha}^J$ denote the eigenvector components in the basis 
$\vert \alpha \vert$. 
Since the definition of $K_l^J$ involves the total number of states $M_J$
which differ for different $J$'s and since the appropriate reference
for our considerations is GOE, we normalise $K_l^J$    
to the GOE limit of this quantity which reads~\cite{Izrail}:
\begin{equation}
K_{GOE}^J = \psi (M_J /2 +1) -\psi(3/2),
\label{eqEGOE}
\end{equation}
where $\psi$ is the digamma function.
Within our model the so-calculated and RQE ensemble averaged quantity
for all the states versus their corresponding energies $E_l^J$ is illustrated
in Fig.~3. As anticipated, it is not $J=0$ whose lowest eigenstate comes
out most localised, i.e., most regular. The lowest states for several higher
$J$ values deviate much more from GOE. This in particular applies to 
$J=2$ and, especially, to $J=4$. This thus indicates more favorable 
conditions for the emergence of energy gaps for larger $J$ than for $J=0$,   
contrary to the numerical outcome of ref.~\cite{Johnson}. 
That our numerical procedure is equivalent to that of ref.~\cite{Johnson}
is confirmed by the fact that we can reproduce the results when
calculating the same quantities.
The question thus arises as how to reconcile such conflicting conclusions.

In fact Fig.~3 provides one hint of what might be the reason.
The $J=0$ states are spread over the broadest energy interval even though
the number of states $(M_0=14)$ is here significantly smaller than for 
several larger $J$ values $(M_1=19, M_2=33, M_3=29, M_4=26)$.
As a result, the average separation between the states is a factor of few
larger for $J=0$ than for the remaining ones. Putting on the states from
different $J$-sectors together, as is done in ref.~\cite{Johnson}, 
is thus likely to hide the genuine character of the effect under consideration.

In order to illustrate the consequence of such a procedure in 
Fig.~4 (dashed line) we show
distributions of the ground state $(E_1^J)$ gaps 
\begin{equation}
s^J=(E_2^J - E_1^J)/D^J, 
\label{eqrho}
\end{equation}
\vskip 0.4cm
\epsfig{file=fig3.eps,width=8cm}
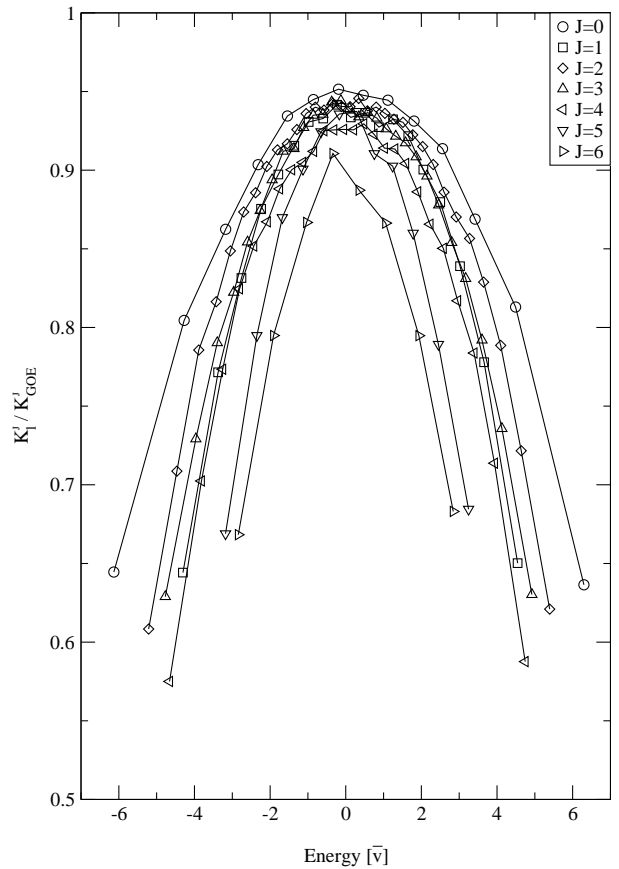
{\bf Fig.~3} The information entropy normalised to its GOE limit
$(K_l^J / K_{GOE}^J)$ for all the states $l$ from various $J$-sectors
(all positive parity)
versus energies $(E_l^J)$ of those states. All the quantities are ensemble
averaged. The energy units are the same as in Fig.~2.\\   
\vskip 0.3cm
where similarly as in ref.~\cite{Johnson}
\begin{equation}
D^J=<E_3^J - E_2^J>, 
\label{eqD}
\end{equation}
for each $J$ individually. 

As it is clearly seen the $J=0$-sector no longer distinguishes significantly
from the remaining ones.
From the point of view of our considerations presented above one would
however expect an even reduced probability for occurance of the large
ground state energy gaps in this particular sector as compared to 
the $J>0$ ones. As the solid lines in Fig.~4 indicate such an effect 
does indeed take place when $D^J$ in eq.~(\ref{eqrho}) is replaced by
\begin{equation}
{\bar D}^J = <E_{M_J}^J - E_2^J>/(M_J-2).
\label{eqDbar}
\end{equation}  
Actually the definition of $D^J$ as specified by eq.~(\ref{eqD}) looks somewhat
arbitrary. It seems more appropriate and more consistent with the above global
considerations to relate the ground state energy gap just to the average
global level spacing among the remaining states, 
characteristic for a given $J$.

Finally one may ask a question why this tendency does not extend to even 
higher $J$-values. In this connection one has to remember that the 
off-diagonal matrix elements is not the only relevant element.
These are the diagonal matrix elements which constitute the driving term.
Irrespective of the value of $J$ their distribution is always Gaussian-like.
This can be observed numerically and is consistent with arguments formulated
in terms of eqs.~(\ref{eqv} - \ref{eqPVul}) since the geometrical factors 
$c^{J \alpha \alpha}_{J' i i'}$ entering the diagonal matrix elements are
always nonnegative. Increasing however $J$ beyond 4 results in a significant 
reduction of the variance of $P_V(v)$ for the diagonal matrix elements 
and consequently a larger fraction of the off-diagonal matrix elements becomes 
effective in mixing the basis states. Thus the effect under investigation
results from an interplay between distributions of the diagonal and 
off-diagonal matrix elements and their relative magnitudes.\\
\vskip 1cm
\epsfig{file=fig4.eps,width=8cm}
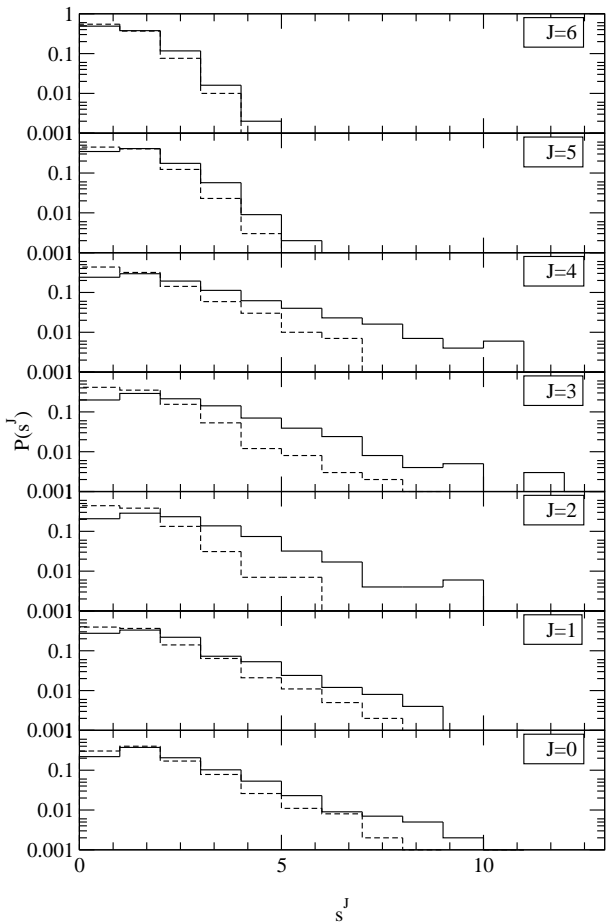
{\bf Fig.~4} Distributions of ground state energy gaps $s^J$ as 
defined by the eq.~(\ref{eqrho}) for successive $J$'s.
The dashed line uses $D^J$ defined by 
eq.~(\ref{eqD}) while the solid line the one defined by eq.~(\ref{eqDbar}).\\
\vskip 0.3cm 

In conclusion, the present investigation based both on theoretical as well
as on numerical arguments clearly shows that the many-body problems 
described in terms of various variants of the two-body random ensembles
(like RQE or TBRE) develop quantitatively well identified deviations from
the GOE. These deviations quantified in terms of localisation or of
energy gaps point to the angular momenta between 2 and 4 as those $J$-sectors
whose ground states are ordered most.     
From this perspective the current interpretation of the numerical results
of ref.~\cite{Johnson} is thus an artifact of mixing states with 
different characteristic energy scales from different $J$-sectors. 
As a side remark it seems also appropriate to notice at this point that the
arguments going in parallel to eqs.~(\ref{eqv}-\ref{eqPVul}) provide a more
adequate approach towards understanding the distribution of matrix elements
in realistic nuclear shell-model calculations than the ones based on 
multipole expansion~\cite{Zelev}.

We acknowledge useful discussions with J.~Oko{\l}owicz, M.~P{\l}oszajczak
and I.~Rotter at the early stage of this development. This work was 
partly supported by KBN Grant No. 2 P03B 097 16.

\end{document}